\documentclass{appolb}
\usepackage{epsfig}
\usepackage{rotate}

\begin{document}


\begin{titlepage}
\begin{flushright}
\begin{tabular}{l}
DESY 00--087\\
hep--ph/0006132\\
June 2000
\end{tabular}
\end{flushright}

\vspace*{1.5truecm}

\begin{center}
\boldmath
{\Large \bf Theoretical Overview of CP Violation\\
\vspace*{0.3truecm}
in $B$-Meson Decay}
\unboldmath

\vspace*{1.2cm}

{\sc{\large Robert Fleischer}}

\vspace*{0.4cm} 

{\it Deutsches Elektronen-Synchrotron DESY, 
Notkestr.\ 85,\\ 
D--22607 Hamburg, Germany}

\vspace{1.3truecm}

{\large\bf Abstract\\[10pt]} \parbox[t]{\textwidth}{
After a brief look at CP violation in kaon decays, a short overview of CP 
violation in the $B$-meson system and of strategies to determine the angles 
of the unitarity triangles of the CKM matrix is given. Both general aspects 
and some recent developments are discussed, including $B_c^\pm\to D^\pm_s D$
and $B\to\pi K$ decays, as well as the $B_d\to\pi^+\pi^-$, $B_s\to K^+K^-$ 
system.}

\vspace{1.5cm}
 
{\sl Invited plenary talk given at the\\
6th International Workshop on Production, Properties\\
and Interaction of Mesons (MESON 2000),\\
Cracow, Poland, 19--23 May 2000\\
To appear in the Proceedings (Acta Physica Polonica B)}
\end{center}

\end{titlepage}
 
\thispagestyle{empty}
\vbox{}
\newpage
 
\setcounter{page}{1}
 

\title{Theoretical Overview of CP Violation in $B$-Meson Decay%
\thanks{Presented at MESON 2000, Cracow, 19--23 May 2000}%
}
\author{Robert Fleischer
\address{Deutsches Elektronen-Synchrotron DESY, 
Notkestr.\ 85,\\ 
D--22607 Hamburg, Germany}
}
\maketitle
\begin{abstract}
After a brief look at CP violation in kaon decays, a short overview of CP 
violation in the $B$-meson system and of strategies to determine the angles 
of the unitarity triangles of the CKM matrix is given. Both general aspects 
and some recent developments are discussed, including $B_c^\pm\to D^\pm_s D$
and $B\to\pi K$ decays, as well as the $B_d\to\pi^+\pi^-$, $B_s\to K^+K^-$ 
system.
\end{abstract}

\section{Introduction}
The non-conservation of the CP symmetry, which was discovered in 1964 in 
neutral kaon decays \cite{CP-obs}, is one of the central aspects of 
modern particle physics, and is still one of the least well experimentally 
constrained phenomena. In particular the $B$-meson system 
provides a very fertile testing ground for the Standard-Model (SM) 
description of CP violation. This feature is also reflected in the tremendous 
effort put in the experimental programmes to explore $B$ physics. 
The BaBar and BELLE detectors are already taking data, HERA-B has seen 
its first events, and CLEO-III, CDF-II and D0-II will follow in the near 
future. Although the physics potential of these experiments is very exciting, 
it may well be that the ``definite'' answer in the search for new physics 
will be left for second-generation $B$-physics experiments at hadron 
machines, such as LHCb or BTeV \cite{LHC}.

\begin{figure}
\begin{tabular}{lr}
   \epsfysize=3.5cm
   \epsffile{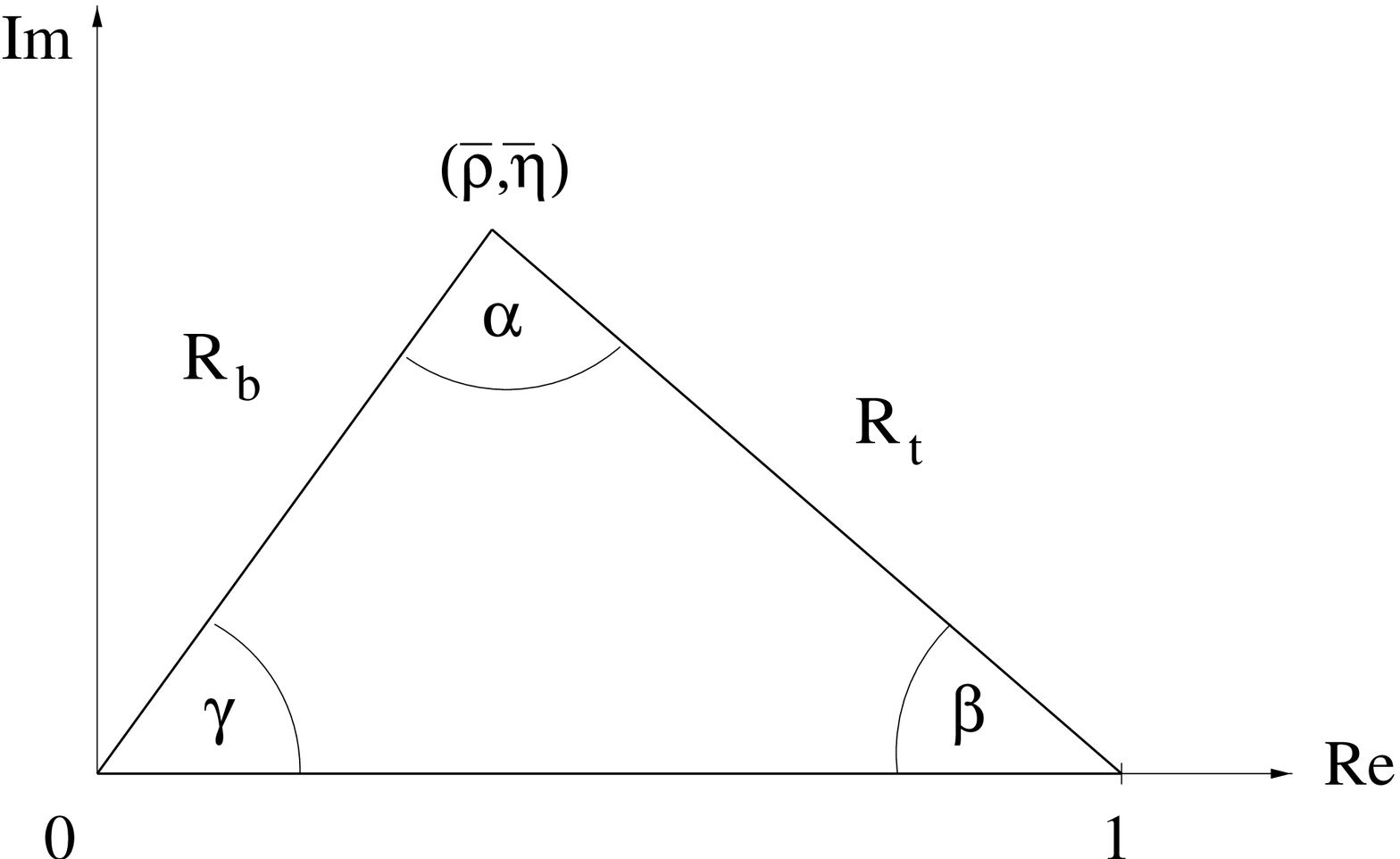}
&
   \epsfysize=3.5cm
   \epsffile{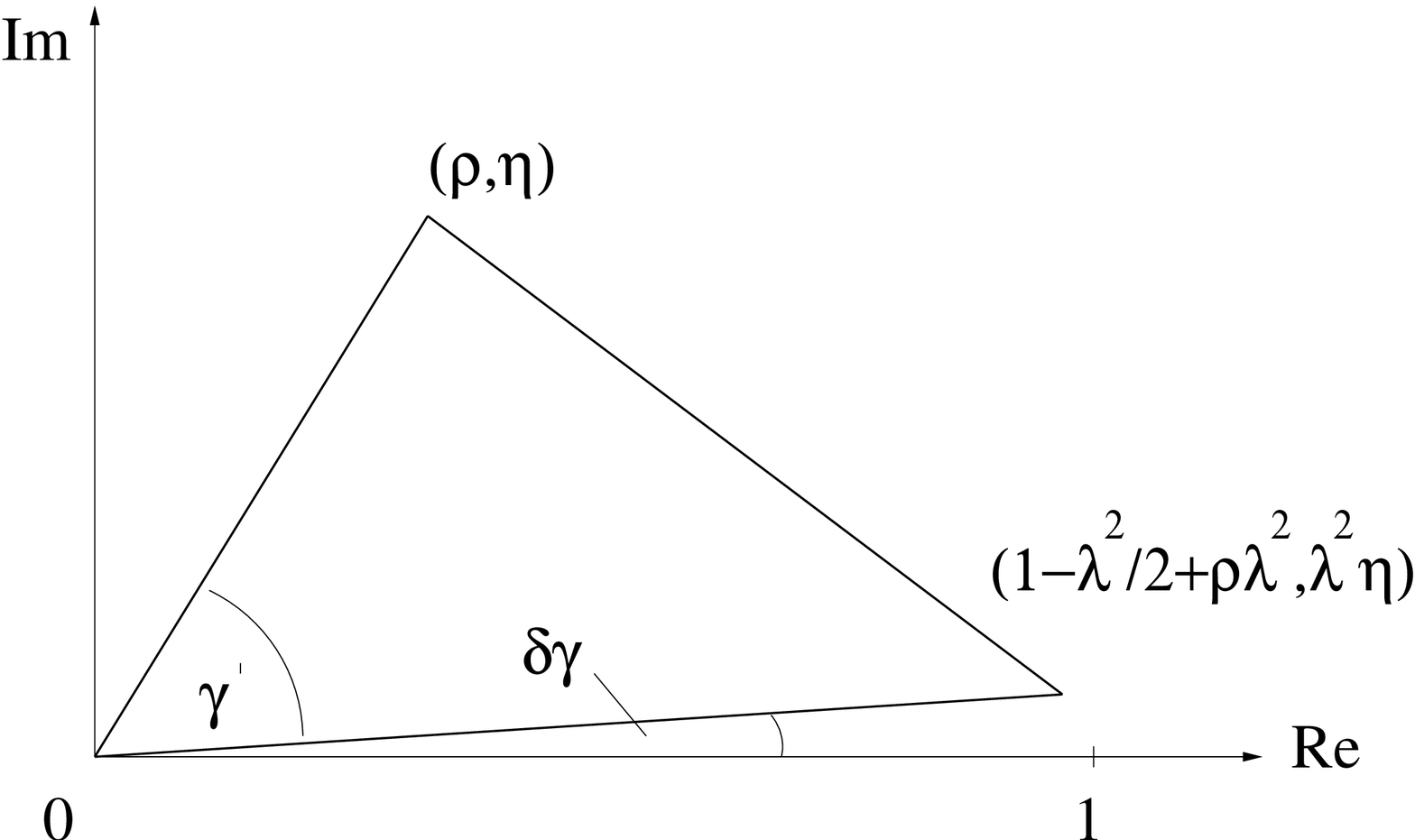}
\end{tabular}
\caption[]{The two non-squashed unitarity triangles of the CKM matrix.
Here, $\overline{\rho}$ and $\overline{\eta}$ are 
related to the Wolfenstein parameters $\rho$ and $\eta$ through 
$\overline{\rho}\equiv\left(1-\lambda^2/2\right)\rho$ and 
$\overline{\eta}\equiv\left(1-\lambda^2/2\right)\eta$, respectively \cite{BLO}.
}
\label{fig:UT}
\end{figure}

Within the framework of the SM, CP violation is closely related 
to the Cabibbo--Kobayashi--Maskawa (CKM) matrix, connecting 
the electroweak eigenstates of the down, strange and bottom quarks with their 
mass eigenstates. As far as CP violation is concerned, the central feature 
is that -- in addition to three generalized Cabibbo-type angles -- also 
a {\it complex phase} is needed in the three-generation case to parametrize 
the CKM matrix. This complex phase is the origin of CP violation within 
the SM. Concerning tests of the CKM picture of CP violation,
the central targets are the unitarity triangles of the CKM matrix. 
The unitarity of the CKM matrix
leads to a set of 12 equations, consisting of 6 normalization and 6 
orthogonality relations. The latter can be represented as 6 triangles
in the complex plane, all having the same area. However, in only 
two of them, all three sides are of comparable magnitude 
${\cal O}(\lambda^3)$, while in the remaining ones, one side is suppressed 
relative to the others by ${\cal O}(\lambda^2)$ or ${\cal O}(\lambda^4)$,
where $\lambda\equiv|V_{us}|=0.22$ denotes the Wolfenstein parameter 
\cite{wolf}. 
The two non-squashed triangles agree at leading order in the Wolfenstein 
expansion (${\cal O}(\lambda^3)$), so that we actually have to 
deal with a single triangle at this order, which is usually referred to as 
``the'' unitarity triangle of the CKM matrix. However, in the era 
of second-generation experiments, starting around 2005, we will have
to take into account the next-to-leading order terms of the Wolfenstein 
expansion, and will have to distinguish between the unitarity triangles 
shown in Fig.~\ref{fig:UT}.

\section{A brief look at the $K$-meson system}
Although the discovery of CP violation goes back to 1964 \cite{CP-obs},
so far this phenomenon has been observed only within the neutral $K$-meson 
system, where it is described by two complex quantities, called $\varepsilon$ 
and $\varepsilon'$, which are defined by the following ratios of decay 
amplitudes:
\begin{equation}\label{defs-eps}
\frac{A(K_{\rm L}\to\pi^+\pi^-)}{A(K_{\rm S}
\to\pi^+\pi^-)}=\varepsilon+\varepsilon',\quad
\frac{A(K_{\rm L}\to\pi^0\pi^0)}{A(K_{\rm S}
\to\pi^0\pi^0)}=\varepsilon-2\,\varepsilon'.
\end{equation}
While $\varepsilon=(2.280\pm0.013)\times e^{i\frac{\pi}{4}}\times 10^{-3}$
parametrizes ``indirect'' CP violation, originating from the fact that
the mass eigenstates of the neutral kaon system are not CP eigenstates,
the quantity Re$(\varepsilon'/\varepsilon)$ measures ``direct'' CP violation 
in $K\to\pi\pi$ transitions. The CP-violating observable $\varepsilon$ plays 
an important role to constrain the unitarity triangle \cite{BF-rev,AL} and 
implies 
in particular a positive value of the Wolfenstein parameter~$\eta$. In 1999,
new measurements of Re$(\varepsilon'/\varepsilon)$ have demonstrated that
this observable is non zero, thereby excluding ``superweak''
models of CP violation \cite{superweak}:
\begin{equation}\label{epsprime-res}
\mbox{Re}(\varepsilon'/\varepsilon)=\left\{\begin{array}{ll}
(28\pm4.1)\times10^{-4}&\mbox{(KTeV Collaboration \cite{KTeV}),}\\
(14\pm4.3)\times10^{-4}&\mbox{(NA48 Collaboration \,\cite{NA48}).}
\end{array}\right.
\end{equation}
Unfortunately, the calculations of Re$(\varepsilon'/\varepsilon)$ are very 
involved and suffer at present from large hadronic uncertainties 
\cite{epsprime}. Consequently, this observable does not allow a powerful 
test of the CP-violating sector of the SM, unless the hadronic 
matrix elements of the relevant operators can be brought under better control. 

In order to test the SM description of CP violation, the 
rare decays $K_{\rm L}\to\pi^0\nu\overline{\nu}$ and 
$K^+\to\pi^+\nu\overline{\nu}$ are more promising and may allow a 
determination of $\sin(2\beta)$ with respectable accuracy \cite{bb}.
Yet it is clear that the kaon system by itself cannot provide the whole 
picture of CP violation, and therefore it is essential to study CP violation 
outside this system. In this respect, $B$-meson decays appear to be most 
promising.
 
\section{The central target: the $B$-meson system}\label{CP-B}

In order to determine the angles of the unitarity triangles shown in
Fig.~\ref{fig:UT}, and to test the SM description of CP 
violation, the major role is played by non-leptonic $B$ decays, which 
can be divided into three decay classes: decays receiving both ``tree'' and 
``penguin'' contributions, pure ``tree'' decays, and pure ``penguin'' 
decays. There are two types of penguin topologies: gluonic (QCD) and 
electroweak (EW) penguins. 
Because of the large top-quark mass, also EW penguins play an important 
role in several non-leptonic $B$-decay processes \cite{rev}. 

\subsection{CP violation in neutral $B$-meson decays}\label{sec:CP-neut}

A particularly simple and interesting situation arises if we restrict 
ourselves to decays of neutral $B_q$-mesons ($q\in\{d,s\}$) into CP 
self-conjugate final states $|f\rangle$, satisfying the relation 
$({\cal CP})|f\rangle=\pm\,|f\rangle$. In this case, the corresponding 
time-dependent CP asymmetry can be expressed as
\begin{eqnarray}
\lefteqn{a_{\rm CP}(t)\equiv\frac{\Gamma(B^0_q(t)\to f)-
\Gamma(\overline{B^0_q}(t)\to f)}{\Gamma(B^0_q(t)\to f)+
\Gamma(\overline{B^0_q}(t)\to f)}=}\nonumber\\
&&2\,e^{-\Gamma_q t}\left[\frac{{\cal A}_{\rm CP}^{\rm dir}(B_q\to f)
\cos(\Delta M_q t)+{\cal A}_{\rm CP}^{\rm mix}(B_q\to f)\sin(\Delta M_q t)}{
e^{-\Gamma_{\rm H}^{(q)}t}+e^{-\Gamma_{\rm L}^{(q)}t}+
{\cal A}_{\rm \Delta\Gamma}(B_q\to f)\left(e^{-\Gamma_{\rm H}^{(q)}t}-
e^{-\Gamma_{\rm L}^{(q)}t}\right)} \right],\label{ee6}
\end{eqnarray}
where $\Delta M_q\equiv M_{\rm H}^{(q)}-M_{\rm L}^{(q)}$ is the
mass difference between the $B_q$ mass eigenstates, and 
the $\Gamma_{\rm H,L}^{(q)}$ denote their decay widths, with
$\Gamma_q\equiv(\Gamma_{\rm H}^{(q)}+\Gamma_{\rm L}^{(q)})/2$.
In Eq.\ (\ref{ee6}), we have separated the ``direct'' from the 
``mixing-induced'' CP-violating contributions, which are described by
\begin{equation}\label{ee7}
{\cal A}^{\mbox{{\scriptsize dir}}}_{\mbox{{\scriptsize CP}}}(B_q\to f)\equiv
\frac{1-\bigl|\xi_f^{(q)}\bigr|^2}{1+\bigl|\xi_f^{(q)}\bigr|^2}\quad
\mbox{and}\quad
{\cal A}^{\mbox{{\scriptsize mix}}}_{\mbox{{\scriptsize
CP}}}(B_q\to f)\equiv\frac{2\,\mbox{Im}\,\xi^{(q)}_f}{1+\bigl|\xi^{(q)}_f
\bigr|^2}\,,
\end{equation} 
respectively. Whereas the width difference 
$\Delta\Gamma_q\equiv\Gamma_{\rm H}^{(q)}-\Gamma_{\rm L}^{(q)}$ is 
negligibly small in the $B_d$ system, it may be 
sizeable in the $B_s$ system (for a recent calculation, 
see \cite{DGamma-cal}), thereby providing the observable 
\begin{equation}\label{ADGam}
{\cal A}_{\rm \Delta\Gamma}(B_q\to f)\equiv
\frac{2\,\mbox{Re}\,\xi^{(q)}_f}{1+\bigl|\xi^{(q)}_f
\bigr|^2}.
\end{equation}
Essentially all the information needed to evaluate the CP asymmetry
(\ref{ee6}) is included in the following quantity \cite{rev}:
\begin{equation}
\xi_f^{(q)}=\mp\,e^{-i\phi_q}\,
\frac{A(\overline{B^0_q}\to f)}{A(B^0_q\to f)}=
\mp\,e^{-i\phi_q}\,
\frac{\sum\limits_{j=u,c}V_{jr}^\ast V_{jb}\,
{\cal M}^{jr}}{\sum\limits_{j=u,c}V_{jr}V_{jb}^\ast\,
{\cal M}^{jr}}\,,
\end{equation}
where the ${\cal M}^{jr}$ denote hadronic matrix elements 
of certain four-quark operators, the label $r\in\{d,s\}$ distinguishes 
between $\overline{b}\to\overline{d}$ and $\overline{b}\to\overline{s}$ 
transitions, and 
\begin{equation}
\phi_q=\left\{\begin{array}{cr}
+2\beta&\mbox{($q=d$)}\\
-2\delta\gamma&\mbox{($q=s$)}\end{array}\right.
\end{equation}
is related to the weak $B_q^0$--$\overline{B_q^0}$ mixing phase. In general, 
the quantity $\xi_f^{(q)}$ suffers from hadronic uncertainties, which are 
due to the hadronic matrix elements ${\cal M}^{jr}$. However, if the decay
$B_q\to f$ is dominated by a single CKM amplitude, the corresponding 
matrix elements cancel, and the convention-independent observable 
$\xi_f^{(q)}$ takes the simple form
\begin{equation}\label{ee10}
\xi_f^{(q)}=\mp\exp\left[-i\left(\phi_q-\phi_{\mbox{{\scriptsize 
D}}}^{(f)}\right)\right],
\end{equation}
where $\phi_{\mbox{{\scriptsize D}}}^{(f)}$ is a weak decay phase, 
which is given by
\begin{equation}
\phi_{\mbox{{\scriptsize D}}}^{(f)}=\left\{\begin{array}{cc}
-2\gamma&\mbox{for dominant 
$\overline{b}\to\overline{u}u\overline{r}$ CKM amplitudes,}\\
0&\,\mbox{for dominant $\overline{b}\to\overline{c}c\overline{r}$ CKM 
amplitudes.}
\end{array}\right.
\end{equation}

\subsubsection{The ``gold-plated'' mode 
$B_d\to J/\psi\, K_{\rm S}$}\label{sec:BdPsiKS}

The most important application of the simple formalism discussed above  
is the decay $B_d\to J/\psi\, K_{\mbox{{\scriptsize S}}}$, which is dominated 
by the $\overline{b}\to\overline{c}c\overline{s}$ CKM amplitude (for a
detailed discussion, see \cite{rev}), implying
\begin{equation}\label{e12}
{\cal A}^{\mbox{{\scriptsize mix}}}_{\mbox{{\scriptsize
CP}}}(B_d\to J/\psi\, K_{\mbox{{\scriptsize S}}})=+\sin[-(2\beta-0)]\,.
\end{equation}
Another non-trivial prediction of the SM is vanishingly small
direct CP violation. Since (\ref{ee10}) applies with excellent accuracy 
to $B_d\to J/\psi\, K_{\mbox{{\scriptsize S}}}$ -- the point is that penguins
enter essentially with the same weak phase as the leading tree
contribution -- it is referred to as the 
``gold-plated'' mode to determine the CKM angle $\beta$ \cite{bisa}.
First attempts to measure $\sin(2\beta)$ through the CP asymmetry 
(\ref{e12}) were already performed \cite{sin2b-exp}:
\begin{equation}
\sin(2\beta)=\left\{\begin{array}{ll}
3.2^{+1.8}_{-2.0}\pm0.5&\mbox{(OPAL Collaboration)}\\
0.79^{+0.41}_{-0.44}&\mbox{(CDF Collaboration)}\\
0.93^{+0.64+0.36}_{-0.88-0.24}&\mbox{(ALEPH Collaboration).}
\end{array}\right.
\end{equation}
Although the experimental uncertainties are still very large, it is 
interesting to note that these results favour the SM expectation of a 
{\it positive} value of $\sin(2\beta)$ \cite{AL}. In the $B$-factory era, an 
experimental uncertainty of $\left.\Delta\sin(2\beta)\right|_{\rm exp}=0.05$ 
seems to be achievable, whereas second-generation experiments of the LHC 
era aim at $\left.\Delta\sin(2\beta)\right|_{\rm exp}={\cal O}(0.005)$
\cite{LHC}. This tremendous experimental accuracy raises the question of 
hadronic uncertainties due to penguin contributions. An interesting channel 
in this context is $B_s\to J/\psi\,K_{\rm S}$, allowing us to control the
-- presumably very small -- 
penguin effects in the determination of
$\phi_d=2\beta$ from $B_d\to J/\psi\,K_{\rm S}$, and to extract the
CKM angle $\gamma$ \cite{RF-Bspsiphi}.

\subsubsection{The decay $B_d\to\pi^+\pi^-$}\label{sec:Bdpipi}

If this mode would not receive penguin contributions, its mixing-induced
CP asymmetry would allow a measurement of $\sin(2\alpha)$:
\begin{equation}
{\cal A}^{\mbox{{\scriptsize mix}}}_{\mbox{{\scriptsize
CP}}}(B_d\to\pi^+\pi^-)=-\sin[-(2\beta+2\gamma)]=-\sin(2\alpha).
\end{equation}
However, this relation is strongly affected by penguin effects, 
which were analysed by many authors \cite{charles}. There are various 
methods to control the corresponding hadronic uncertainties. 
Unfortunately, these strategies are usually rather challenging from 
an experimental point of view. 

The best known approach was proposed by Gronau and London \cite{gl}. 
It makes use of an $SU(2)$ isospin relation between the $B^+\to\pi^+\pi^0$,
$B^0_d\to\pi^+\pi^-$ and $B^0_d\to\pi^+\pi^-$ decay amplitudes, as
well as their CP conjugates,  
which can be represented as two triangles in the complex plane.
Unfortunately, the Gronau--London approach suffers from a serious
experimental problem, since the measurement of $\mbox{BR}(B_d\to\pi^0\pi^0)$
is very difficult. 

Alternative methods to control the penguin uncertainties in the 
extraction of $\alpha$ from $B_d\to\pi^+\pi^-$ are very desirable. An 
important one for the $e^+$--$e^-$ $B$-factories is provided 
by $B\to\rho\,\pi$ modes \cite{Brhopi}. Here the isospin triangle relations 
are replaced by pentagonal relations, and the corresponding approach is 
rather complicated. As we will see in more detail below, an interesting
strategy for hadron machines to employ the CP-violating observables 
of $B_d\to\pi^+\pi^-$ is offered by $B_s\to K^+K^-$, allowing a 
simultaneous determination of $\beta$ and $\gamma$ without any 
penguin uncertainties \cite{BsKK}.

\subsubsection{The $B_s$-meson system}\label{sec:Bs}

Since the $e^+$--\,$e^-$ $B$-factories operating at $\Upsilon(4S)$ 
are not in a position to explore the $B_s$ system, it
is of particular interest for hadron machines. There are important
differences to the $B_d$ system: the $B^0_s$--$\overline{B^0_s}$ 
mixing phase $\phi_s=-2\lambda^2\eta={\cal O}(0.03)$
is negligibly small in the SM, and a large mixing parameter
$x_s\equiv\Delta M_s/\Gamma_s={\cal O}(20)$ is expected. Moreover, the
expected sizeable width difference $\Delta\Gamma_s$ provides interesting
strategies to extract CKM phases from ``untagged'' $B_s$ data samples,
where the rapid oscillating $\Delta M_st$ terms cancel \cite{untagged}. 
Among the $B_s$ benchmark modes are $B_s\to D_s^\pm K^\mp$, allowing
a theoretically clean determination of the CKM phase $\gamma-2\delta
\gamma$ \cite{ADK}, and $B_s\to J/\psi\, \phi$. This decay offers
interesting strategies to extract $\Delta M_s$, $\Delta\Gamma_s$ and
$\phi_s$ from the angular distribution of the 
$J/\psi[\to l^+l^-]\,\phi[\to K^+K^-]$ decay products \cite{DDFF}. 
Since $B_s\to J/\psi\, \phi$ modes exhibit, moreover, very small 
CP-violating effects in the SM, they represent an interesting probe 
for new-physics contributions to $B^0_s$--$\overline{B^0_s}$ mixing 
\cite{nir-sil,BaFl}.

\subsection{CP violation in charged $B$-meson decays}
Since there are no mixing effects present in charged $B$-meson decays, 
non-vanishing CP asymmetries ${\cal A}_{\mbox{{\scriptsize CP}}}$
would give us unambiguous evidence for ``direct'' CP violation in the 
$B$ system. Such CP asymmetries arise from the interference between 
decay amplitudes with both different CP-violating weak and different 
CP-conserving strong phases. In the SM, the weak phases are 
related to the phases of the CKM matrix, whereas the strong phases 
are induced by FSI processes. In general, the strong 
phases introduce severe theoretical uncertainties into the calculation of 
${\cal A}_{\mbox{{\scriptsize CP}}}$, thereby destroying the clean 
relation to the CP-violating weak phases. 

An important tool to overcome these problems is provided by amplitude 
relations between certain non-leptonic $B$ decays. The prototype of this 
approach, which is due to Gronau and Wyler \cite{gw}, uses 
$B^\pm \to K^\pm D$ decays. If the
$D$-meson is observed as a CP eigenstates, amplitude triangles can be
construced, allowing a theoretically clean determination of $\gamma$.
Unfortunately, these triangles turned out to be highly stretched,
and are -- from an experimental point of view -- not very useful to 
determine $\gamma$. Further difficulties were pointed out in \cite{ads}.
As an alternative, the decays $B_d\to K^{\ast 0}D$ were proposed \cite{dun} 
because the triangles are more equilateral. But all sides are small 
because of various suppression mechanisms. In another paper, the triangle 
approach to extract $\gamma$ \cite{gw} was also extended to the $B_c$ 
system \cite{masetti}. At first sight, here everything is completely 
analogous to $B^\pm_u\to K^\pm D$. However, there is an important difference 
\cite{fw}: in the $B_c^\pm\to D_s^\pm D$ system, the amplitude with the 
rather small CKM matrix element $V_{ub}$ is not colour suppressed, while 
the larger element $V_{cb}$ comes with a colour-suppression factor. 
Therefore, the two amplitudes are similar in size! In contrast to this 
favourable situation, in the $B^\pm_u\to K^\pm D$ system, the matrix 
element $V_{ub}$ comes with the colour suppression factor, resulting 
in a very stretched triangle, while in the decays $B_d\to K^{*0} D$, all 
amplitudes are colour suppressed. Decays of the type $B_c^\pm \to D^\pm D$ 
-- the $U$-spin counterparts of $B_c^\pm\to D_s^\pm D$ -- can be added to 
the analysis, as well as channels, where the $D_s^\pm$- and $D^\pm$-mesons
are replaced by higher resonances. At the LHC, one expects about $10^{10}$ 
untriggered $B_c$\,s per year of running. Provided there are no serious 
experimental problems, the $B_c^\pm\to D_{(s)}^\pm D$ approach should be 
very interesting for the corresponding $B$-physics programme. 

\subsection{Probing $\gamma$ with $B\to\pi K$ decays}\label{sec:bpiK}
In order to obtain direct information on $\gamma$, $B\to\pi K$ decays 
are very promising \cite{BpiK}, and have received a lot of attention 
during the recent years \cite{BpiK-revs}. Because of the small ratio 
$|V_{us}V_{ub}^\ast/(V_{ts}V_{tb}^\ast)|\approx0.02$, these modes are 
dominated by penguin topologies and are hence very sensitive to 
new-physics contributions \cite{NP}. Interestingly, already CP-averaged 
$B\to\pi K$ branching ratios may imply highly non-trivial constraints 
on $\gamma$ \cite{FM}. So far, the studies of these bounds have focussed 
on the following two systems: $B_d\to\pi^\mp K^\pm$, $B^\pm\to\pi^\pm K$ 
\cite{FM}, and $B^\pm\to\pi^0K^\pm$, $B^\pm\to\pi^\pm K$ \cite{NR}. 
Recently, it was pointed out that also the neutral decays 
$B_d\to\pi^\mp K^\pm$ and $B_d\to\pi^0 K$ may be very interesting in this 
respect \cite{BF,BF2}. 

The $B\to\pi K$ strategies to probe $\gamma$ make use of flavour-symmetry 
arguments ($SU(2)$ or $SU(3)$), and rely, in addition, on dynamical 
assumptions, concerning mainly the smallness of certain rescattering 
processes, such as $B^+\to\{\pi^0K^+\}\to\pi^+K^0$. The theoretical 
understanding of such FSI processes is poor at present \cite{BBNS}. However, 
there are important experimental indicators for possible large rescattering 
effects, e.g.\ $B^+\to K^+\overline{K^0}$ or $B_d\to K^+K^-$, and methods to 
include them in the strategies to probe $\gamma$. 

In order to constrain $\gamma$ through $B\to\pi K$ decays, the key 
quantities are ratios $R_{({\rm c,n})}$ of CP-averaged branching ratios, 
which can be constructed for the ``mixed'', charged and neutral $B\to\pi K$ 
systems listed above.
Employing the theoretical ingredients sketched in the previous paragraph, 
we obtain
\begin{equation}
R_{({\rm c,n})}=R_{({\rm c,n})}(\gamma,q_{({\rm c,n})},r_{({\rm c,n})},
\delta_{({\rm c,n})}),
\end{equation}
where $q_{({\rm c,n})}$ denotes the ratio of EW penguins to trees, 
$r_{({\rm c,n})}$ is the ratio of trees to QCD penguins, and 
$\delta_{({\rm c,n})}$ is the CP-conserving strong phase between 
tree and QCD penguin amplitudes. Whereas $q_{({\rm c,n})}$ can be fixed 
through theory, and $r_{({\rm c,n})}$ with the help of additional experimental 
information, e.g.\ on BR$(B^\pm\to\pi^\pm\pi^0)$, $\delta_{({\rm c,n})}$ 
suffers from large hadronic uncertainties and is essentially unknown. 
However, we can get rid of $\delta_{({\rm c,n})}$ by keeping it as a 
``free'' variable, yielding minimal and maximal values for $R_{({\rm c,n})}$:
\begin{equation}\label{const1}
\left.R^{\rm ext}_{({\rm c,n})}\right|_{\delta_{({\rm c,n})}}=
\mbox{function}(\gamma,q_{({\rm c,n})},r_{({\rm c,n})}).
\end{equation}
Keeping in addition $r_{({\rm c,n})}$ as a free variable, we obtain 
another -- less restrictive -- minimal value for $R_{({\rm c,n})}$:
\begin{equation}\label{const2}
\left.R^{\rm min}_{({\rm c,n})}\right|_{r_{({\rm c,n})},\delta_{({\rm c,n})}}
=\kappa(\gamma,q_{({\rm c,n})})\sin^2\gamma.
\end{equation}
Since values of $\gamma$ corresponding to $R^{\rm exp}_{({\rm c,n})}<
R^{\rm min}_{({\rm c,n})}$ or
$R^{\rm exp}_{({\rm c,n})}>R^{\rm max}_{({\rm c,n})}$, where  
$R^{\rm exp}_{({\rm c,n})}$ denotes the measured value
of $R_{({\rm c,n})}$, are excluded, (\ref{const1}) and (\ref{const2}) 
imply an allowed range for $\gamma$. Although it is too early to draw
definite conclusions, it is interesting to note that the most recent 
CLEO results on $R_{({\rm c,n})}$ are in favour of strong constraints on
$\gamma$, where the second quadrant, i.e.\ $\gamma\geq 90^\circ$,
is preferred. Such a situation would be in conflict with the standard 
analysis of the unitarity triangle, yielding 
$38^\circ\leq\gamma\leq81^\circ$ \cite{AL}.

The observables  $R_{({\rm c,n})}$ imply also constraints on 
$\delta_{({\rm c,n})}$, where the present CLEO data are in favour of 
$\cos\delta_{\rm c}>0$ and $\cos\delta_{\rm n}<0$, which would 
be in conflict with the theoretical expectation of equal signs for 
$\cos\delta_{\rm c}$ and $\cos\delta_{\rm n}$~\cite{BF2}. If future 
data should confirm this ``puzzle'', it may be an indication for new-physics 
contributions to the EW penguin sector, or a manifestation of 
large non-factorizable $SU(3)$-breaking effects. In order to distinguish 
between these possibilties, detailed studies of the various patterns of 
new-physics effects in all $B\to\pi K$ decays are essential, as well as 
critical analyses of possible sources for $SU(3)$ breaking.
As soon as CP asymmetries ${\cal A}^{({\rm c,n})}_{\rm CP}$ in 
$B_d\to\pi^\mp K^\pm$ or $B^\pm\to\pi^0K^\pm$ are observed, we may go 
beyond the bounds and may determine $\gamma$ and $\delta_{({\rm c,n})}$. 
The physics potential of $B\to\pi K$ decays is very interesting and plays 
a central role for the $B$-factories.
 
\subsection{Extracting $\beta$ and $\gamma$ from $B_d\to\pi^+\pi^-$ and
$B_s\to K^+K^-$}

There are interesting strategies to extract CKM phases with the help of
$U$-spin-related $B$ decays, where all down and strange quarks are
interchanged with each other \cite{snowmass}. A particularly interesting 
one is provided by the decays $B_d\to\pi^+\pi^-$ and $B_s\to K^+K^-$, 
allowing a simultaneous determination of $\beta$ and $\gamma$ \cite{BsKK}. 
This new strategy is not affected by any penguin topologies -- it rather 
makes use of them -- and does not rely on certain ``plausible'' dynamical or 
model-dependent assumptions. Moreover, FSI effects, which led to considerable 
attention in the context of the determination of $\gamma$ from $B\to\pi K$ 
decays, as we have noted in Subsection~\ref{sec:bpiK}, do not lead to any 
problems. The theoretical accuracy is only limited by $U$-spin-breaking 
effects, which vanish in the factorization approximation in the present 
case. This strategy is ideally suited for LHCb 
($\Delta\gamma={\cal O}(1^\circ)$) \cite{LHC}, and is also 
very promising for CDF-II \cite{wuerthwein}. Conceptually similar approaches
are provided by $B_{s(d)}\to J/\psi\,K_{\rm S}$ or $B_{d(s)}\to 
D^+_{d(s)}D^-_{d(s)}$ decays \cite{RF-Bspsiphi}.

\vspace*{0.5truecm}

\section{Conclusions and outlook}

The phenomenology of non-leptonic $B$ decays is very rich and provides 
a fertile testing ground for the SM description of CP violation. As a 
by-product, interesting insights into hadronic physics can be 
obtained. There is no doubt that an exciting future -- the $B$-physics 
era of particle physics -- is ahead of us. Hopefully, it will shed 
light on the physics beyond the SM.

\end{document}